\pdfoutput=1
\documentclass[a4paper,article]{memoir}

\usepackage[]{textcomp}
\usepackage[]{microtype}
\usepackage[utf8]{inputenc}
\usepackage[T1]{fontenc}
\usepackage[]{hyperref}
\usepackage[]{graphicx}
\usepackage[]{amsthm}
\usepackage[]{amssymb}

\usepackage{hyperref}

\input{style.tex}

\newtheorem{thm__4}{Lemma}\newtheorem{thm__3}{Remark}\newtheorem{thm__2}{Theorem}\newtheorem{thm__1}{Definition}

\begin{document}
  \title{Circuits via topoi}\author{Arnaud Spiwack}{\maketitle}
  \begin{abstract}
    Leveraging topos theory a semantics can be given to sequential circuits where time-sensitive gates, such as unit delay, are treated uniformly with combinational gates. Both kinds of gates are functions in a particular topos: the topos of presheaves over the natural ordering of $\mathbb{N}$. This is used to show that sequential circuits validate the equational theory of traced categories.
  \end{abstract}
  When giving semantics to circuits (typically boolean circuits), it is customary to treat the combinational~--~\emph{i.e.} time-independent~--~parts of the circuits differently from time sensitive ones. Since it is usually assumed that the only time-sensitive gate is the unit delay, each outgoing wire from a delay is considered an additional input, and each incoming wire an additional output. Some care is taken to feed the right output into the right input at next iteration, and so time-sensitivity is eliminated and one can reason on a purely combinational circuit.
  \par
  This is not very convenient to reason equationally about moving unit delays for better placement. But this approach really breaks down when considering time-sensitive gates which are not simple unit delay. This article takes its root in the study of compilation of the circuit programming language Faust~\cite{Orlarey2004}. Faust features a somewhat unusual kind of delay gate, written \textsf{s\symbol{64}d}, where \textsf{s} is an arbitrary signal, and \textsf{d} is a \emph{time-varying} bounded natural number signal whose value at time $t$ determines how far in the past of \textsf{s} to fetch the value of the delayed signal.
  \par
  With such a construct, it becomes impossible to ``cut'' a circuit into a combinational circuit. At least not without heavy modifications (the constraint that \textsf{d} is bounded is imposed in order to be able to compile the program in constant memory, so such a circuit can be reduced to use only unit delays).
  \par
  To address this issue, let us turn to presheaves and topos theory. The critical property which we shall use is that topoi are models of constructive mathematics. Therefore we shall first develop a theory of combinational circuits in ordinary constructive mathematics, then lift it to sequential circuits \emph{via} a presheaf construction.
  \par
  Much work has been put~\cite{Bonchi2015,Baez2015}, recently, in using category theory to explain and exploit the linear algebraic aspects of circuits from control-theory. This article explores an orthogonal axis of the design space. Both can, and should, in principle be combined to obtain linear algebra with time. It is what signal processing is made of.
  \par
  Before we move on, I have to start with an apology: despite the subject of topoi and presheaves being rather technical, I will be assuming quite a bit of familiarity with them in this article. I realise that this will make this article unnecessarily arduous for many. But in order for this article to be written at all, I felt I had to limit its scope so. A good, exhaustive, introduction to topos theory can be found in Mac Lane \& Moerdijk's \emph{Sheaves in geometry and logic}~\cite{MacLaneMoerdijk1992}.
  \par
  \paragraph{Acknowledgement}
  I was once sitting with Noam Zeilberger listening to a seminar by Gérard Berry. About circuits. Berry's presentation was obviously of great interest to the both of us as we went on discussing its content for quite a while after that. At some point Zeilberger remarked: ``I don't really know what a circuit \emph{is}'', and I suddenly realised that I didn't quite either; despite the intuitive, and occasionally concrete, nature of circuits. If Zeilberger's remark gave you pause, as it did to me, then read on for my attempt at a definition.
  \chapter{Combinational circuits}\label{latex_lib_label_1}
  \par
  It is direct to give a definition of combinational circuits if they are not allowed to have loops: just interpret each gate as a function a compose things appropriately. Or, more generally, if you are so disposed, interpret each gate as an arrow in some cartesian category, and interpret appropriately.
  \par
  The case that drove my interest, however, requires loops~--~aka feedback. I am not particularly in need of delay-free loops, although this is of legitimate interest (see for instance~\cite{Mendler2012}), but the semantics which is developed in this section will be lifted in Section~\ref{latex_lib_label_6} to sequential circuits which are useless without some form of feedback.
  \par
  \section{Constructive domain semantics}
  \par
  As good computer scientist ought to when faced with tricky fixed point (even in circuits, loops are, after all, fixed points), let us turn to domains. Before we give a formal description of our semantics, let me note that it is a straightforward variant of the rather venerable \emph{three-valued semantics} of combinational boolean circuits~\cite{Malik1993}, which, by the way, has been shown, with \emph{caveats}, to be a good semantics for electronic circuits~\cite{Mendler2012}.
  \par
  Boolean circuits have, of course, a special relevance in computing science due to their being the basic building block of computers. But we will not restrict ourselves so. Wires will be allowed to carry values of any type we wish. The Faust programming language, for instance, has wires of type $\mathbb{N}$ and $\mathbb{R}$ (floating point numbers, in practice). The types, which, for the purpose of this article, are simply sets of permitted values, allowed for the wires by a circuit language will be called \emph{base types}.
  \par
  \begin{thm__1}[Bounded height domain]
    \label{latex_lib_label_3}A bounded height domain is a partially ordered set $D$ equipped with a number $b$ such that every increasing chain ${x}_{1}\leqslant \mbox{\ldots }\leqslant {x}_{n}$ in $D$ with $n>b$ has a pair ${x}_{i}\geqslant {x}_{j}$ with $i<j$ (equivalently, for every $i\leqslant k\leqslant j$, ${x}_{i}={x}_{k}$).
  \end{thm__1}
  \par
  Circuits will be given a semantics as increasing functions between such domains. To the extent that the material present in this section is different from the usual treatment it is to render this section constructive to be compatible with the topos of Section~\ref{latex_lib_label_6}. This is the reason why we focus on bounded height domains rather than the more usual $\omega $-\textsc{cpo}s. Note also that $b$, in Definition~\ref{latex_lib_label_3} is \emph{not} the height of the domain but rather an upper bound on this height. The reason is that, constructively, there may not be an exact height (see also~\cite{CoquandSpiwack2010} for more thoughts on finiteness in constructive mathematics). Every proof, in this section, is constructive.
  \par
  Bounded height domain have the fixed-point property, just like other kinds of domains. Note that the fixed-point property of \textsc{cpo}s or $\omega $-\textsc{cpo}s are also constructive. The added value of bounded height domains is that there are really few constructive \textsc{cpo}s or $\omega $-\textsc{cpo}s (see Remark~\ref{latex_lib_label_5} below). Another practical advantage of bounded height domain is that the fixed-point property applies to all increasing functions, which will free us from proving continuity.
  \par
  \begin{thm__2}[Fixed-point property]
    \label{latex_lib_label_4}Every increasing function $f : D\rightarrow D$ for a bounded height domain $D$ with a smallest element $\bot $ has a smallest fixed-point.
  \end{thm__2}
  \begin{proof}
    Let $b$ be a bound on the height of $D$. The sequence $\bot \leqslant f{\left( \bot \right) }\leqslant {f}^{2}{\left( \bot \right) }\leqslant \mbox{\ldots }\leqslant {f}^{b}{\left( \bot \right) }$ has length $b+1$. By definition, there is an $i<b$ such that ${f}^{i}{\left( \bot \right) }={f}^{i+1}{\left( \bot \right) }$ hence ${f}^{b}{\left( \bot \right) }={f}^{i}{\left( \bot \right) }$ is a fixed point.
    \par
      It is the smallest since, by induction, for any fixed point ${x}_{0}$ of $f$ and every $k$, ${f}^{k}{\left( \bot \right) }\leqslant {x}_{0}$.  
  \end{proof}
  \par
  Remark that we can refine the proof to show that the least fixed point of $f$ is also its least pre-fixed point (\emph{i.e.} such that $f{\left( x\right) }\leqslant x$).
  \par
  Base types, which are sets, can be naturally identified to bounded height domains (without a smallest element).
  \par
  \begin{thm__1}[Flat domains]
    Given a set $A$, the partially ordered set (also noted $A$) where $x\leqslant y\iff x=y$ is a domain of height bounded by $1$, which we call a flat domain.
  \end{thm__1}
  \par
  Continuing on the subject of constructiveness, notice that flat domains are an example of domain which can't be assigned a height. Indeed, if $A$ is inhabited then $A$ has height $1$, whereas when $A$ is empty then $A$ has height $0$ but it is not possible, in constructive mathematics to decide whether $A$ is empty or not, the height flat domains is, therefore, not well defined (it is, in fact, impossible to define a non-constant integer-valued function on sets~\cite{Escardo2013}).
  \par
  In order to use base types in conjunction with the fixed point property they need a smallest element which we add freely.
  \par
  \begin{thm__1}[Lifted domains]
    Given a domain $A$ with a bound $b$ on its height, we construct a domain ${A}_{\bot }$, called \emph{lifted}, by adding a distinguished element $\bot $ to $A$ and considering it smaller than every element of $A$: $\forall {x}^{\in A}.\,\,\bot <x$. The height of ${A}_{\bot }$ is bounded by $b+1$
  \end{thm__1}
  \par
  Wires in circuits will be interpreted as taking value in the lifted flat domains corresponding to base types. Increasing functions between lifted flat domain are such that if $f{\left( \bot \right) }\ne \bot $, then for any $a$, $f{\left( a\right) }=f{\left( \bot \right) }$. In particular, if $f$ is such a function with several fixed points, then $f{\left( \bot \right) }=\bot $ and the smallest fixed point is $\bot $. So $\bot $ represents both the absence of a well-defined fixed point and the presence of several fixed points.
  \par
  Obviously, in order for the smallest fixed point of$f$ to be non-$\bot $, $f$ needs to ignore some of its input wires, for instance $f$ could be the well-know parallel or:
  \par
  $
  \begin{array}{c|ccc}
    \mathsf{por} & \bot  & 0 & 1\\
    \midrule
    \bot  & \bot  & \bot  & 1\\
    0 & \bot  & 0 & 1\\
    1 & 1 & 1 & 1\\
  \end{array}
  $
  \par
  When one of the input of the parallel or is $1$, then the output is $1$, whatever the behaviour of the other input. In particular the following circuit is well defined (it outputs $1$):
  \par
  \begin{displaymath}
    \makebox[1.\linewidth]{\makebox[0.1\linewidth]{}\parbox{0.9\linewidth}
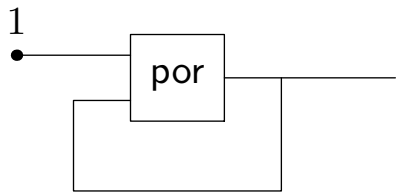
{\includegraphics{topocircuit-melt-figure1.mps}}}
  \end{displaymath}
  More useful examples can be found in~\cite{Malik1993,Mendler2012}.
  \par
  \begin{thm__3}
    \label{latex_lib_label_5}Lifted flat domains are an example of bounded-height domains which are not necessarily ${\omega}$-\textsc{cpo}s, constructively. Indeed consider $\mathbf{1}=\left\{ 0\right\} $, the singleton set, then ${\mathbf{1}}_{\bot }$ is not constructively an ${\omega}$-\textsc{cpo}. An ${\omega}$-chain in ${\mathbf{1}}_{\bot }$ is an infinite sequence of $\bot $ and $0$ (such that after a $0$, every element is $0$). If an ${\omega}$-chain has $\bot $ as an upper bound, then all of its elements are $\bot $, if an ${\omega}$-chain has $0$ as an upper bound, at least one of its elements must be $0$\footnote{Accomplished constructive mathematicians may noticed that I have made use of Markov's principle in this statement: it's a valid thing to do, though since if something is unprovable from Markov's principle, it is certainly unprovable without. Alternatively, ``must'' in that sentence can be interpreted as the double-negation modality, in which case the statement is constructively true, and leads to a weaker, still non-constructive, version of the limited principle of omniscience.}. If every ${\omega}$-chain had an upper bound, it would give a way to decide whether they contain a $0$ or not, which is equivalent to the limited principle of omniscience: a known-to-be-non-constructive principle.
  \end{thm__3}
  \par
  To formalise circuits with multiple wires we remark that bounded height domains are closed by cartesian products.
  \par
  \begin{thm__4}[Cartesian product of domains]
    The product $A\times B$ of two bounded height domains of respective bound ${b}_{A}$ and ${b}_{B}$, with order $\left( {x}_{1},{y}_{1}\right) \leqslant \left( {x}_{2},{y}_{2}\right) \iff {x}_{1}\leqslant {x}_{2}\land {y}_{1}\leqslant {y}_{2}$ , is a domain whose height is bounded by ${b}_{A}\times {b}_{B}$.
  \end{thm__4}
  \begin{proof}
    A chain in $A\times B$ is simply a list of pairs $\left( x,y\right) $, with an constraint on consecutive pairs. However, chains can be represented differently as a list of pairs $\left( x,l\right) $ with $l$ a chain in $B$, with the intent that the pair $\left( x,\left[ {y}_{1},\mbox{\ldots },{y}_{n}\right] \right) $ represents the chain $\left( x,{y}_{1}\right) \leqslant \mbox{\ldots }\leqslant \left( x,{y}_{n}\right) $. So that chains now have multiple representations depending on how successive pairs with the same first component are ``contracted''.
    \par
      We begin with the simplest representation where every pair $\left( x,l\right) $ is an $\left( x,\left[ y\right] \right) $. Now, as long as our list is of size longer than ${b}_{A}$ (without loss of generality we can suppose both ${b}_{A}$ and ${b}_{B}$ to be non-zero) we can use the fact that the first components describe a chain to find consecutive positions with the same $x$ which we can contract. Hence strictly reducing the size of our list. This process gives us a contracted representation of the long chain of length ${b}_{A}$ or less. But, since the total length of the chain is larger than ${b}_{A}\times {b}_{B}$, there must be at least one second-component list with length larger than ${b}_{B}$. Applying the definition of the bound ${b}_{B}$ to this list concludes the proof.
  \end{proof}
  \par
  Combinational gates are, therefore, interpreted as increasing functions of type ${{A}_{1}}_{\bot }\times \mbox{\ldots }\times {{A}_{n}}_{\bot }\rightarrow {{A}_{n+1}}_{\bot }\times \mbox{\ldots }\times {{A}_{p}}_{\bot }$ (with each ${A}_{i}$ being a base type). 
  \par
  \section{Traced category}\label{latex_lib_label_2}
  \par
  What is left is to use the fixed-point property to make precise the definition of feedback wires, the solution is given by Hasegawa~\cite[Theorem 3.1]{Hasegawa1997} who gives a method to transform fixed-point operators into traces.
  \par
  Traced categories~\cite{Joyal1996} provide a graphical language which is essentially the same as circuits with feedback. It is reassuring that circuits can be interpreted as arrows in such a traced category. It provides a natural equational theory on circuits which can be, among other things, leveraged to produce optimisation schemes~\cite{Liu2009}.
  \par
  \begin{thm__4}[Local fixed-point property]
    Theorem~\ref{latex_lib_label_4} can be extended to produce a local fixed point function: let $f : A\times X\rightarrow X$ an increasing function ($A$ and $X$ bounded height domain with a smallest element), there is an increasing function $\mu {\left( f\right) } : A\rightarrow X$, such that for any $a : A$, $\mu {\left( f\right) }{\left( a\right) }$ is the least fixed point of the increasing function $\lambda x.\,f{\left( a,x\right) }$.
  \end{thm__4}
  \begin{proof}
    The proof bulk of the proof is the same as Theorem \ref{latex_lib_label_4}, taking into account that, by definition of cartesian product $\lambda x.\,f{\left( a,x\right) }$ is, indeed, increasing: let $b$ a bound on the height of X,
      $\mu {\left( f\right) }{\left( a\right) }={\left( \lambda x.\,f{\left( a,x\right) }\right) }^{b}{\left( \bot \right) }$.
    \par
    We need to check the $\mu {\left( f\right) }$ is indeed increasing. But since $f$ is increasing, for any $a\leqslant a'$ and any $x\leqslant x'$  $f{\left( a,x\right) }\leqslant f{\left( a',x'\right) }$; by induction, we conclude that $\mu {\left( f\right) }{\left( a\right) }\leqslant \mu {\left( f\right) }{\left( a'\right) }$.
  \end{proof}
  \par
  Hasegawa tells us that there are three properties to verify for a fixed point operator to yield a trace (note that, reciprocally, all traces in a cartesian category yield such a fixed point operator). We shall write ${\mu }_{{a}}x.\,\,f{\left( a,x\right) }$ instead of $\mu {\left( \lambda \left( a,x\right) .\,f{\left( a,x\right) }\right) }$. In addition and by definition, $\left( {\mu }_{{a}}x.\,\,f{\left( a,x\right) }\right) {\left( {a}_{0}\right) }={\mu }_{{}}x.\,\,f{\left( {a}_{0},x\right) }$; since the former is cumbersome, we will use the latter as a shorthand.
  \par
  \begin{thm__4}[Naturality in $A$]
    For any $f : A\times X\rightarrow X$ and $g : B\rightarrow A$, the following holds: ${\mu }_{{b}}x.\,\,f{\left( g{\left( b\right) },x\right) }=\mu {\left( f\right) }\circ g$.
  \end{thm__4}
  \begin{proof}
    Let $b : B$, $\left( {\mu }_{{b}}x.\,\,f{\left( g{\left( b\right) },x\right) }\right) {\left( b\right) }={\mu }_{{}}x.\,\,f{\left( g{\left( b\right) },x\right) }$ is, by definition, the least fixed point of $\lambda x.\,f{\left( g{\left( b\right) },x\right) }$. And, also by definition, so is $\mu {\left( f\right) }{\left( g{\left( b\right) }\right) }$.
  \end{proof}
  \par
  \begin{thm__4}[Naturality in $X$]
    Let $f$ be an increasing function in $A\times X\rightarrow Y$. For any $g : Y\rightarrow X$, ${\mu }_{{a}}x.\,\,g{\left( f{\left( a,x\right) }\right) }=g\circ \left( {\mu }_{{a}}y.\,\,f{\left( a,g{\left( y\right) }\right) }\right) $.
  \end{thm__4}
  \begin{proof}
    Let us fix an $a : A$.
      \begin{itemize}[{\tiny{${\blacksquare}$}}]
      \item Let us prove that $g{\left( {\mu }_{{}}y.\,\,f{\left( a,g{\left( y\right) }\right) }\right) }$ is a fixed point of $\lambda x.\,g{\left( f{\left( a,x\right) }\right) }$, and therefore ${\mu }_{{}}x.\,\,g{\left( f{\left( a,x\right) }\right) }\leqslant g{\left( {\mu }_{{}}y.\,\,f{\left( a,g{\left( y\right) }\right) }\right) }$. This follows immediately from the fact that $f{\left( a,g{\left( {\mu }_{{}}y.\,\,f{\left( a,g{\left( y\right) }\right) }\right) }\right) }={\mu }_{{}}y.\,\,f{\left( a,g{\left( y\right) },a\right) }$ and the fact that $g$ is increasing.
      \item Conversely, we prove similarly that $f{\left( a,{\mu }_{{}}x.\,\,g{\left( f{\left( a,x\right) }\right) }\right) }$ is a fixed point of $\lambda y.\,f{\left( a,g{\left( y\right) }\right) }$. This yields $f{\left( a,{\mu }_{{}}x.\,\,g{\left( f{\left( a,x\right) }\right) }\right) }\geqslant {\mu }_{{}}y.\,\,f{\left( a,g{\left( y\right) }\right) }=f{\left( a,g{\left( {\mu }_{{}}y.\,\,f{\left( a,g{\left( y\right) }\right) }\right) }\right) }$, and then, ${\mu }_{{}}x.\,\,g{\left( f{\left( a,x\right) }\right) }\geqslant g{\left( {\mu }_{{}}y.\,\,f{\left( a,g{\left( y\right) }\right) }\right) }$ by monotonicity of $\lambda x.\,f{\left( a,x\right) }$.
    \end{itemize}
    \par
      The two inequalities prove the lemma.
  \end{proof}
  \par
  \begin{thm__4}[Bekič]
    Let $f : A\times X\times Y\rightarrow X$ and $g : A\times X\times Y\rightarrow Y$. Taking $h : A\rightarrow X$ to be $h{\left( a\right) }={\mu }_{{}}x.\,\,f{\left( a,x,\mu {\left( g\right) }{\left( a,x\right) }\right) }$, the following holds ${\mu }_{{a}}\left( x,y\right) .\,\,\left( f{\left( a,x,y\right) },g{\left( a,x,y\right) }\right) =\lambda a.\,\left( h{\left( a\right) },\mu {\left( g\right) }{\left( a,h{\left( a\right) }\right) }\right) $
  \end{thm__4}
  \begin{proof}
    For $a : A$, let us prove that $\left( h{\left( a\right) },\mu {\left( g\right) }{\left( a,h{\left( a\right) }\right) }\right) $ is a fixed point of $\lambda \left( x,y\right) .\,\left( f{\left( a,x,y\right) },g{\left( a,x,y\right) }\right) $.
      \begin{displaymath}
      \begin{array}{lll}
         & \left( f{\left( a,h{\left( a\right) },\mu {\left( g\right) }{\left( a,h{\left( a\right) }\right) }\right) },g{\left( a,h{\left( a\right) },\mu {\left( g\right) }{\left( a,h{\left( a\right) }\right) }\right) }\right)  & \\
        = & \left( f{\left( a,h{\left( a\right) },\mu {\left( g\right) }{\left( a,h{\left( a\right) }\right) }\right) },\mu {\left( g\right) }{\left( a,h{\left( a\right) }\right) }\right)  & \mbox{(definition of $\mu {\left( g\right) }$)}\\
        = & \left( h{\left( a\right) },\mu {\left( g\right) }{\left( a,h{\left( a\right) }\right) }\right)  & \mbox{(definition of $h$)}\\
      \end{array}
    \end{displaymath}
    \par
      We also have that
      \begin{displaymath}
      \begin{array}{ll}
         & {\mu }_{{}}\left( x,y\right) .\,\,\left( f{\left( a,x,y\right) },g{\left( a,x,y\right) }\right) \\
        = & \left( f{\left( a,{\mu }_{{}}\left( x,y\right) .\,\,\left( f{\left( a,x,y\right) },g{\left( a,x,y\right) }\right) \right) },g{\left( a,{\mu }_{{}}\left( x,y\right) .\,\,\left( f{\left( a,x,y\right) },g{\left( a,x,y\right) }\right) \right) }\right) \\
      \end{array}
    \end{displaymath}
    \par
      This allows us to test both components for being fixed points or the corresponding function, which will suffice to conclude.
    \par
      \begin{itemize}[{\tiny{${\blacksquare}$}}]
      \item $f{\left( a,{\mu }_{{}}\left( x,y\right) .\,\,\left( f{\left( a,x,y\right) },g{\left( a,x,y\right) }\right) \right) }\geqslant h{\left( a\right) }$: by definition of $h$ it suffices to show that $f{\left( a,{\mu }_{{}}\left( x,y\right) .\,\,\left( f{\left( a,x,y\right) },g{\left( a,x,y\right) }\right) \right) }$ is a pre-fixed point of $\lambda x.\,f{\left( a,x,\mu {\left( g\right) }{\left( \left( a,x\right) \right) }\right) }$. After tedious calculations\footnote{So tedious, in fact, that I ended up formalising most of this section in the Coq proof assistant which, contrary to me, is not susceptible to calculation mistakes. Plus, I was getting lost and could use the help. This goes to prove that for certain mathematical proofs, proof assistant can be a productive way to develop proofs.}, it amounts to proving, calling $\left( {x}_{0},{y}_{0}\right) ={\mu }_{{}}\left( x,y\right) .\,\,\left( f{\left( a,x,y\right) },g{\left( a,x,y\right) }\right) $, that ${\mu }_{{}}y.\,\,g{\left( a,{x}_{0},y\right) }\leqslant {y}_{0}$. It is easily checked that ${y}_{0}$ is a fixed point of $\lambda y.\,g{\left( a,{x}_{0},y\right) }$, which concludes this sub-proof.
      \item $g{\left( a,{\mu }_{{}}\left( x,y\right) .\,\,\left( f{\left( a,x,y\right) },g{\left( a,x,y\right) }\right) \right) }\geqslant \mu {\left( g\right) }{\left( a,h{\left( a\right) }\right) }$. The argument is similar to above.
    \end{itemize}
      \end{proof}
  \chapter{Sequential circuits}\label{latex_lib_label_6}
  \par
  Adding time-sensitive gates forces to change the semantics. Sequential circuits are not to be seen as functions from (product of) base types to base types, but rather as functions from streams of base types to stream of base types. Unfortunately, the type ${A}^{\mathbb{N}}$ of streams of a finite height domain $A$ is not a finite height domain in any useful way.
  \par
  To be able to model feedback, a change a perspective will be needed. The typical approach to analysis of sequential circuits with feedback is to ``cut'' unit delays making their incoming wire into a special new output and their outgoing wire into a special new input. What makes this transformation even meaningful is the requirement that to compute a finite prefix of length $n$ of a circuit's output, only a finite prefix of length $n$ of the input is necessary. This requirement is called \emph{causality}.
  \par
  \section{Causal sets}
  \par
  With that in mind, it makes sense to see streams not as a whole, but as a progression of prefixes ${\left( {A}^{n}\right) }_{n\in \mathbb{N}}$. All of the ${A}^{n}$, by virtue of being finite products of finite height domains, are finite height domains. A causal function can, then, be defined as a collection ${\left( {f}_{n}\right) }_{n\in \mathbb{N}}$ of functions ${A}^{n}\rightarrow {B}^{n}$ such that ${f}_{n+1}{\left( w\cdot a\right) }$ is of the form ${f}_{n}{\left( w\right) }\cdot b$.
  \par
  To abstract over these notions, let us introduce a topos~--~\emph{i.e.} a model of constructive mathematics~--~where such a presentation of streams and causal functions is natural.
  \par
  \begin{thm__1}[Causal sets]
    The topos of \emph{causal sets} is the topos of presheaves over the set of natural number with its standard ordering.
  \end{thm__1}
  \par
  This topos has been extensively studied by Birkedal, Møgelberg, Schwinghammer \& Støvring~\cite{DBLP:journals/corr/abs-1208-3596} under the name \emph{topos of trees} to contribute to the related problem of step-indexing. Their article can serve as a reference.
  \par
  A causal set is, therefore, given by a family ${\left( {A}_{n}\right) }_{n\in \mathbb{N}}$ of sets together with \emph{restriction} functions ${r}_{n} : {A}_{n+1}\rightarrow {A}_{n}$. Causal functions are families of functions ${\left( {f}_{n}\right) }_{n\in \mathbb{N}}$ such that ${r}_{n}{\left( {f}_{n+1}{\left( a\right) }\right) }={f}_{n}{\left( {r}_{n}{\left( a\right) }\right) }$. Streams, seen, as above, as a progression of prefixes, form a causal set ${\mathbb{S}}_{A}$ with ${\left( {\mathbb{S}}_{A}\right) }_{n}={A}^{n}$ and ${r}_{n}{\left( w\cdot a\right) }=w$. Causal functions, in the sense of the topos of causal set, on ${\mathbb{S}}_{A}$ are the same as causal arrows of streams; so that arrows in the topos of causal sets are, indeed, a generalisation of causal functions of streams.
  \par
  By analogy with streams, the sets ${A}_{n}$, constituting the causal set $A$, are called the sets of prefixes of $A$ or just prefixes of $A$. Since the indices of the functions can often be inferred from the context, they will often be omitted; for example: the compatibility of $f$ with restrictions may be written $r{\left( f{\left( a\right) }\right) }=f{\left( r{\left( a\right) }\right) }$.
  \par
  Topos are models of constructive mathematics, hence there is an interplay between \emph{internal} statements of the topos of causal sets which are derived using the rules of constructive mathematics and \emph{external} statements of ordinary mathematics. Internal statements are related to external statements via the Kripke-Joyal semantics~\cite[Section VI.6]{MacLaneMoerdijk1992}: when $\varphi $ is an internal proposition in context $\Gamma $ ($\Gamma $ is a (conjunction of) causal set giving sense to the free variables of $\varphi $), then for $n\in \mathbb{N}$ and ${\alpha }_{n}\in {\Gamma }_{n}$ an external proposition $n\models \varphi {\left( {\alpha }_{n}\right) }$ is defined. The proposition $n\models \varphi {\left( {\alpha }_{n}\right) }$ means that $\varphi $ holds at least until and including time $n$ on ${\alpha }_{n}$. The main property being that if $\varphi $ is provable in constructive mathematics (usually written $\vdash \varphi $), then for all $n$ and ${\alpha }_{n}$, $n\models \varphi {\left( {\alpha }_{n}\right) }$. And conversely, if $\varphi $ is such that $n\models \varphi {\left( {\alpha }_{n}\right) }$ then $\varphi $ is internally valid.
  \par
  \section{Causal domains}
  \par
  Let us now endeavour to give an external description of internal bounded height domain, so as to show that ${\mathbb{S}}_{{A}_{\bot }}$ is an internal bounded height domain for some base type $A$.
  \par
  The ordering relation is reflexive: $\vdash x\leqslant x$. That is, $n\models {\alpha }_{n}\leqslant {\alpha }_{n}$ for any $n$ and ${\alpha }_{n}$. In other words, a reflexive causal relation, is a family of reflexive relations on each set of prefixes (compatible with restrictions). The same holds for symmetry and transitivity, such that an internal ordering relation is an ordering relation on each prefix\footnote{The reader may be worried about the implication in the statement of symmetry and transitivity, since the interpretation of implication is not direct in the Kripke-Joyal semantics. But it doesn't matter at ``toplevel'': $\vdash x\leqslant y\rightarrow y\leqslant x$ translates to $\forall n,{\alpha }_{n},{\beta }_{n}.\,\,\forall k\leqslant n.\,\,k\models {r}^{n-k}{\left( {\alpha }_{n}\right) }\leqslant {r}^{n-k}{\left( {\beta }_{n}\right) }\rightarrow k\models {r}^{n-k}{\left( {\beta }_{n}\right) }\leqslant {r}^{n-k}{\left( {\alpha }_{n}\right) }$ which is equivalent to $\forall n,{\alpha }_{n},{\beta }_{n}.\,\,n\models {\alpha }_{n}\leqslant {\beta }_{n}\rightarrow n\models {\beta }_{n}\leqslant {\alpha }_{n}$.}.
  \par
  The translation of ordering relations illustrate the purpose of causal sets: to make it possible to reason on finite prefixes of infinite data. This is, indeed, what we were looking for, to be able to use the fact that prefixes of ${\mathbb{S}}_{{A}_{\bot }}$ are bounded height domains. We should expect, at this point, that internal bounded height domains are exactly those causal sets where ${A}_{n}$ is a bounded height domain for each $n$, which is indeed the case.
  \par
  The key observation is that the casual $\mathsf{List}{\left( A\right) }$ which is the initial algebra of the functor $A\times X+1$ can be defined as ${\left( \mathsf{List}{\left( A\right) }\right) }_{n}=\mathsf{List}{\left( {A}_{n}\right) }$. The restriction functions act pointwise on the elements of each list. Therefore, since subsets are taken pointwise \emph{i.e.} ${\left\{ x\,{\in }\,A\mid \varphi {\left( x\right) }\right\} }_{n}=\left\{ x\,{\in }\,{A}_{n}\mid {\varphi }_{n}{\left( x\right) }\right\} $, chains internal to the topos of causal sets are chains on prefixes (compatible with restrictions).
  \par
  Thanks to this observation, the internal definition of bounded height domain can be interpreted: ${b}_{n}\in \mathbb{N}$ is a bound on the height of $A$ at time $n$ if for all $k\leqslant n$, ${b}_{n}$ is a bound on the height of ${A}_{k}$ in the ordinary sense. Since being a bound is a monotonous property on $b$, an internal bounded-height domain is a causal set with all prefixes being externally bounded-height domains. An internal domain $A$ has a smallest element if each of the ${A}_{n}$ has and restrictions map smallest elements to smallest elements.
  \par
  \begin{thm__4}
    The causal set ${\mathbb{S}}_{{A}_{\bot }}$, for some ordinary set $A$, is a finite height domain with a smallest element internal to the topos of causal sets.
  \end{thm__4}
  \begin{proof}
    Since the height ${\left( {\mathbb{S}}_{{A}_{\bot }}\right) }_{n}={A}_{\bot }^{n}$ is bounded by ${2}^{n}$, which also bounds all the ${A}_{\bot }^{k}$ for $k\leqslant n$. The smallest element of ${A}_{n}$ is $\left( \bot ,\mbox{\ldots },\bot \right) $.
  \end{proof}
  \par
  As a consequence, we can build circuits as causal increasing functions ${\mathbb{S}}_{{{A}_{1}}_{\bot }}\times \mbox{\ldots }\times {\mathbb{S}}_{{{A}_{n}}_{\bot }}\rightarrow {\mathbb{S}}_{{{A}_{n+1}}_{\bot }}\times \mbox{\ldots }\times {\mathbb{S}}_{{{A}_{p}}_{\bot }}$ and feedback wires can be interpreted as internal least fixed point like in Section~\ref{latex_lib_label_1}. The $n$-th prefix of a causal increasing function is a sequence ${\left( {f}_{i}\in {A}_{i}\rightarrow {B}_{i}\right) }_{i\leqslant n}$ each of the ${f}_{i}$ being increasing, and such that ${r}_{i}{\left( {f}_{i+1}{\left( a\right) }\right) }={f}_{i}{\left( {r}_{i}{\left( a\right) }\right) }$.
  \par
  What remains to be figured out is what a fixed point internal to the topos of causal set is. The internal formula for a fixed point is $f{\left( a\right) }=a$ which translates to ${f}_{n}{\left( {a}_{n}\right) }={a}_{n}$ for any $n$ (note that $f$, being an internal function, \emph{i.e.} an element of ${A}^{A}$, has prefixes ${\left( {f}_{i}\in {A}_{i}\rightarrow {A}_{i}\right) }_{i\leqslant n}$, so ${f}_{n}\in {A}_{n}\rightarrow {A}_{n}$). Therefore $a$ is an internal fixed point if and only if it is a fixed point at each prefix.
  \par
  \section{Lifting traces}
  \par
  The results of the above section, while elegant, do not demonstrate effectively the usefulness of the topos-theoretic semantics: indeed, the treatment of the previous section could have been carried out directly just as easily without requiring topos-theoretic baggage. However, when all this material is developed, it becomes possible to easily lift more powerful theorems directly from the combinational semantics. Let us apply this principle to Hasegawa's theorem from Section~\ref{latex_lib_label_2}.
  \par
  Hasegawa's theorem being an external statement about categories, we will have to translate the statement (but, crucially, not the proofs) of all four lemmas and show that they correspond to the hypotheses of Hasegawa's theorem. Fortunately, this is rendered easy thanks to some standard properties: causal functions are the same as global sections $1\rightarrow {B}^{A}$ (where $1$ is the terminal causal set: ${1}_{n}$ is the singleton set for every $n$), internal equality is interpreted as external equality, and the terms $\lambda x.\,x$ and $\lambda x.\,f{\left( g{\left( x\right) }\right) }$ are interpreted as identity and composition, respectively. From these, we can immediatly deduce that it makes sense to speak of a causal function which is internally increasing, and therefore, that the causal bounded-height domains with a smallest elements and internally increasing causal functions form a subcategory of the topos of causal sets.
  \par
  A slightly trickier property is the local fixed-point operator whose existence is proven internally. Remember that $\exists {x}^{\in A}.\,\,P{\left( x\right) }$ is interpreted by $\forall {n}^{\in \mathbb{N}}.\,\,\exists {{x}_{n}}^{\in {A}_{n}}.\,\,P{\left( {x}_{n}\right) }$ there are no connection between the ${x}_{n}$ chosen at each $n$ so there is not necessarily a global section $1\rightarrow A$ that witnesses the existential. However, when the $x$ not only exists, but is also unique, then, since $P{\left( {x}_{n+1}\right) }\Rightarrow P{\left( {r}_{n}{\left( {x}_{n+1}\right) }\right) }$, then the ${x}_{n}$ necessarily respect the restriction maps of $A$, hence form a global section. There is, of course, at most one local least fixed-point map, hence, internal existence guarantees external existence of a global section, which can be turned into an external map $\mu $ from causal functions to causal functions.
  \par
  The rest of the properties: that $\mu $ is, indeed, a local fixed-point map, that it is natural in $A$ and $X$ and that it verifies Bekič's lemma, are all universally quantified equalities involving composition of arrows (and $\mu $). They can be changed into their categorical counterparts with just a bit of fiddling.
  \par
  We can, therefore, conclude, with barely any proof pertaining to time, that sequential circuits obey the laws of traced categories.
  \chapter*{Conclusion}
  \par
  The topos theoretic approach to the theory of sequential circuits could be unfolded and give rise to a semantics free of all things toposes. As I have hinted in the course of this article, the semantics itself would not be particularly complex, however proofs are significantly simpler when making use of the internal logic of the topos (sometimes called the synthetic point of view). Proving that sequential circuits form a traced category, for instance, almost completely ignored the difference between sequential and combinational circuits.
  \par
  This article can be seen both as a contribution to the growing body of applications of the synthetic approach to mathematical problems, and as a proposal to further the understanding of the mathematics of circuits which, despite being a fundamental concept in computer science is still rather obscure and hard to reason about.
  \par
  I should mention that the circuits considered are partial, in that they may return $\bot $, an ill-formed value. We are really interested in total circuits which do not; but as is the case for total recursive functions, total circuits have no reason to be composable. Fortunately, it is reasonably easy to characterise total circuits: there is a (natural) causal function ${\eta }_{A} : {\mathbb{S}}_{A}\rightarrow {\mathbb{S}}_{{A}_{\bot }}$ we say that a circuit $c : {\mathbb{S}}_{{A}_{\bot }}\rightarrow {\mathbb{S}}_{{B}_{\bot }}$ is total if the composite function $c\circ {\eta }_{A}$ factors through ${\eta }_{B}$, \emph{i.e.} if there is $c' : {\mathbb{S}}_{A}\rightarrow {\mathbb{S}}_{B}$ such that $c\circ {\eta }_{A}={\eta }_{B}\circ c'$. This is straightforwardly extended to several inputs and outputs.
  \par
  Circuits are built by composition and taking a trace. Composition of total circuits is total, only taking a trace can turn a total circuit into a non-total one. The standard way to take a trace safely is to ensure that ``somewhere on the path'' there is a delay. This condition is captured, in the topos of causal sets, by the notion of contractivity which can be internalised and used synthetically~\cite{DBLP:journals/corr/abs-1208-3596}. Contractivity is an example of notion, in the internal logic of the topos of causal sets, which goes beyond standard constructive mathematics. It, indeed, ensures that trace can be taken safely.
  \par
  What may render the synthetic approach difficult is, beyond the need to use constructive mathematics, is the translation of a synthetic statement into an ordinary one. It is tedious and precise, and, though it probably gets better with training, it is hard to convince oneself no error has been made in the process. This is where a proof assistant would be of great help, and be much more efficient at such a task than a human. There is a prototype for the Coq proof assistant by Jaber, Sozeau \& Tabareau~\cite{Jaber2012} which handles the special case of presheaves over a preorder. It is sufficient for the topos in this article, so I could have used it to help with the translations of Section~\ref{latex_lib_label_6}. Since most of Section~\ref{latex_lib_label_1} has already been formalised in Coq\footnote{The formalisation can be found at the following address: \url{https://gist.github.com/aspiwack/628761dab886728bf4db}}, it is not particularly far-fetched. However, I have unfortunately not taken time to learn how to use this tool.
  \par
  To conclude, I feel I should say a few words about syntax, after spending this article on the semantics of circuits. When working with circuits we tend to assume that reorganising of wires preservers syntax so that the following two composition of diagonals are equal:
  \begin{displaymath}
    \makebox[1.\linewidth]{\makebox[0.1\linewidth]{}\parbox{0.9\linewidth}
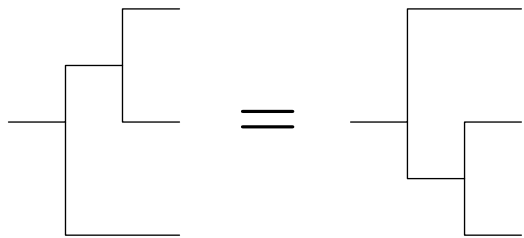
{\includegraphics{topocircuit-melt-figure2.mps}}}
  \end{displaymath}
  But that rearranging gates does not:
  \begin{displaymath}
    \makebox[1.\linewidth]{\makebox[0.1\linewidth]{}\parbox{0.9\linewidth}
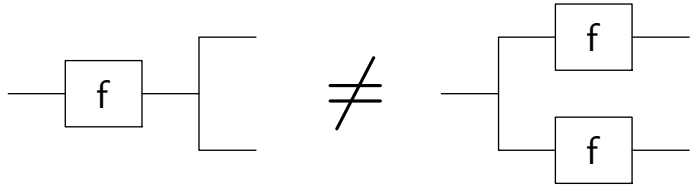
{\includegraphics{topocircuit-melt-figure3.mps}}}
  \end{displaymath}
  \par
  It is customary to take syntax to be a free something, and since our semantics is a traced cartesian category, we may be tempted to take the syntax of circuits to be the free traced cartesian category but that would identify both sides in the latter diagram. Instead, the syntax of circuits should be the free construction of some kind of traced categories with diagonals, where diagonals and augmentations (wires to the empty product) have the usual co-associativity and co-neutrality laws, but are not natural transformations.\bibliography{library}
\end{document}